\documentclass[aps, pra, superscriptaddress, twocolumn, amsfonts, amsmath, amssymb, floatfix]{revtex4-2}
\usepackage{graphicx}
\usepackage{float}
\usepackage{sidecap}
\usepackage{dcolumn}
\usepackage{bm}
\usepackage{epsfig}
\usepackage{braket}
\usepackage{color}
\usepackage{ulem}
\usepackage{amsmath}
\usepackage{wasysym}
\usepackage[colorlinks=true, letterpaper=true, pdfstartview=FitV, linkcolor=blue, citecolor=blue, urlcolor=blue]{hyperref}

\begin{document}    

\title{Fulde-Ferrell superfluids in an asymmetric three-component Fermi Gas}

\author{Yuhan Lu}

\affiliation{Institute for Quantum Science and Technology, Department of Physics, Shanghai University, Shanghai 200444, China}

\author{Lihong Zhou}
\email{lihongzh@shu.edu.cn}
\affiliation{Institute for Quantum Science and Technology, Shanghai University, Shanghai 200444, China}

\author{Yongping Zhang}
\email{yongping11@t.shu.edu.cn}
\affiliation{Institute for Quantum Science and Technology, Department of Physics, Shanghai University, Shanghai 200444, China}

\begin{abstract}
An asymmetric three-component Fermi gas, featuring Raman-induced spin-orbit coupling between the first and second components and contact interaction only between the first and third components, introduces both spin-orbit coupling and population imbalance-two mechanisms known to stabilize the Fulde-Ferrell superfluids.We systematically study Fulde-Ferrell superfluids in an asymmetric three-component  Fermi gas { in two dimensions and at zero temperature} by finding the global minima of the thermodynamic potential. We reveal a new class of composite Fulde-Ferrell superfluids that emerges when strong spin-orbit coupling generates a double-well structure in momentum space within the lower spin-orbit-coupled band. The key features of these composite superfluids are identified.

\end{abstract}

\maketitle

\section{Introduction}\label{introduction}

The degenerate Fermi gases, experimentally realized in ultracold neutral atoms, provide a fundamental platform for investigating paring superfluidity~\cite{RevModPhys.80.885, schrieffer2018theory, giorgini2008theory,schumacher2026observation}. Owing to their parameter tunability, these systems also enable access to the unitary regime and the so-called BCS–BEC crossover~\cite{strinati2018bcs,regal2004observation, zwierlein2004condensation,  randeria1990superconductivity, marini1998evolution}. 


Among exotic pairing superfluids in Fermi gases,  Fulde-Ferrell (FF) pairs~\cite{fulde1964superconductivity}, characterized by an order parameter with a nonzero pairing momentum, and Larkin-Ovchinnikov (LO) pairs~\cite{larkin1965nonuniform}, featured by a spatially modulated order parameter, have attracted considerable research attention~\cite{casalbuoni2004inhomogeneous,radzihovsky2010imbalanced,xu2015topological,yi2015pairing}.  In a multiple-component Fermi gas, population imbalance induces a mismatch between the Fermi surfaces of different components~\cite{zhang2005pairing,sheehy2006bec,ozawa2010population,paananen2006pairing, de2009three,zwierlein2006fermionic, partridge2006pairing,parish2007finite, liao2010spin,tuzemen2025hierarchy}.
Such Fermi surface mismatch arising from different energy bands has become an important mechanism for stabilizing FF superfluids. With the experimental advancement of synthetic spin-orbit coupling (SOC) in ultracold atoms~\cite{lin2011spin,wang2012spin, cheuk2012spin}, SOC has been recognized as another mechanism for the emergence of FF pairs~\cite{qu2013topological,zhang2013topological}. The combination of SOC and an in-plane Zeeman field gives rise to an asymmetric Fermi surface with respect to the center of momentum within  a single spin-orbit-coupled energy band. FF pairing occurs between states on this asymmetric surface.  Meanwhile, SOC may also introduce topological properties to quasiparticles. Spin-orbit-coupled FF superfluids have been intensively investigated in two-component~\cite{yi2012molecule,zhou2011topological,wu2013unconventional, chen2013inhomogeneous, liu2013topological, zheng2013route,   zhou2013exotic, iskin2013topological,dong2013fulde,xu2014anisotropic, cao2014gapless, zheng2014fflo, zheng2020cavity, morales2024resonance} and three-component~\cite{qin2015three, qiu2016universal,chen2014three, zheng2024synthetic} systems.   Moreover, SOC brings new physics to the BCS-BEC crossover~\cite{gong2011bcs,hu2011probing,yu2011spin,he2012bcs,chen2012bcs}.

Ref.~\cite{zhou2014three} proposed a new mechanism of FF pairing formation in an asymmetric three-component Fermi gas, which consists of SOC between the first and second components and contact interactions between the first and third components. This mechanism incorporates both SOC and population imbalance together and can generate two kinds of FF pairs originating from two spin-orbit-coupled energy bands~\cite{zhou2014three}. Since there is no conventional s-wave Cooper pair in this system, FF superfluids can exist in a broad parameter region.   

In this paper, we systematically study FF superfluids in such the asymmetric three-component Fermi gas. { The system is two dimensional and is at zero temperature.}  The SOC, which exists only between the first and second components, causes these two components to share the same chemical potential  $\mu_{12}$ and gives rise to two spin-orbit-coupled energy bands. In contrast, the third component, characterized by a chemical potential  $\mu_3$, forms an isolated energy band. Due to the asymmetry of the interactions, pairing can only occur between the first and third components. 
If $\mu_{12}$ intersects with both spin-orbit-coupled energy bands, we identify two distinct types of FF superfluids, between which a first-order phase transition can be induced by varying  $\mu_3$. On the other hand, if $\mu_{12}$
intersects only with the lower spin-orbit-coupled band, only one type of FF superfluid emerges. These findings are consistent with the results reported in Ref.~\cite{zhou2014three}.

If the SOC dominates, the lower spin-orbit-coupled band exhibits a double-well structure in momentum space.  Letting $\mu_2$ intersect this double-well structure can give rise to two states dominated by the first component. These two states combine to form a composite first-component state, whose average momentum corresponds to one of the minima of the double-well. Meanwhile, $\mu_3$ intersects with the isolated band, generating the other composite state-dominated by the third component-with an average momentum of zero. Ultimately, these two composite states bind together to form a FF pair. The most notable feature of such a composite FF superfluid is that its pairing momentum remains invariant even when  $\mu_{12}$ and $\mu_3$ vary. This newly identified class of composite FF superfluids not only completes the study presented in Ref.~\cite{zhou2014three}, but also introduces a novel mechanism for the composite formation of FF pairs.

The rest of this paper is organized as follows. In Sec.~\ref{Model}, we present the mean-field theoretical framework for the asymmetric three-component Fermi gas. In Sec.~\ref{WeakSEC}, FF superfluids are investigated for the weak SOC case. In Sec.~\ref{StrongSEC}, we reveal a new composite FF superfluid for the strong SOC case. Finally, a conclusion is given in Sec.~\ref{Conclusion}.

\section{Model}
\label{Model}

 We consider three-component Fermi gases confined in a quasi-two-dimension potential~\cite{schumacher2026observation,zhou2014three}. The system is described by the Hamiltonian,
\begin{equation}
H = \sum_{\bm{k}} \Phi^\dagger(\bm{k}) H_0(\bm{k})  \Phi(\bm{k}) + H_\text{I},
\label{Hamiltonian}
 \end{equation}
 where $\Phi(k)=(
c_{\bm{k},1},  c_{\bm{k},2},  c_{\bm{k},3} )^T$ is the wave function with $ c_{\bm{k},i}$  being annihilation operator in momentum space $\bm{k}=(k_x,k_y)$ for the $i$-th component.      
$H_0$ is the single-particle Hamiltonian in momentum space,
\begin{equation}
H_0(\bm{k}) =
\begin{pmatrix}
\varepsilon_{\bm{k}} - \mu_{12} + \alpha k_x & h & 0 \\
h & \varepsilon_{\bm{k}} - \mu_{12} - \alpha k_x & 0 \\
0 & 0 & \varepsilon_{\bm{k}} - \mu_3
\end{pmatrix}.
\label{SingleParticleH}
\end{equation}
Here, 
\begin{equation}
\varepsilon_{\bm {k}}=\frac{\hbar^2\bm{k}^2}{2m}, 
    \end{equation}
is the single-particle kinetic energy with $m$ being atom mass. The first and second components are resonantly coupled via a two-photon transition induced by a pair of Raman lasers~\cite{wang2012spin, cheuk2012spin}. $h$ is the two-photon coupling strength. Since the two Raman beams propagating oppositely in the $x$ direction, the two-photon transition is accompanied by SOC with the strength $\alpha$ along the $\hat{k}_x$ direction. The SOC manifests itself as the presence of $\pm \alpha k_x$ in the first and second component respectively. Furthermore, the Raman coupling gives rise to the same chemical potential $\mu_{12}$ to the first two components. While, the third component does not have a coupling with the previous ones, therefore, it has an independent chemical potential $\mu_3$.  
The interacting  $H_\text{I}$ in Eq.~\eqref{Hamiltonian} describes the contact interaction between the first and third components,
\begin{equation}
\begin{aligned}
   H_\text{I}=\frac{U_{13}}{S}\sum_{\bm{q},\bm{k},\bm{k'}}\left(c_{\bm{k},3}^{\dagger}c_{\bm{q}-\bm{k},1}^{\dagger}c_{\bm{q}-\bm{k'},1}c_{\bm{k'},3}\right),
    \end{aligned}
    \label{eq3}
\end{equation}
 where $U_{13}$ is the interaction rate and $S$ is the area of two dimensional gas. The momentum conservation during collision generates three momenta $ (\bm{k}, \bm{k'}, \bm{q}) $. 
 The rate $U_{13}$ in two-dimension can be renormalized as
\begin{equation}
    \frac{S}{U_{13}}=-\sum_{\bm {k}}\frac{1}{2\varepsilon_{\bm{k}}+E_{b,13}},
    \label{eq4}
\end{equation}
by introducing the binding energy
$E_{b,13}$ of the two-body bound state in the first and third components~\cite{randeria1990superconductivity, marini1998evolution}.

The Hamiltonian in Eq.~\eqref{Hamiltonian} describes an asymmetric three-component Fermi system: (i) The SOC only exists in the first two components, leading them to share the same chemical potential 
 $\mu_{12}$; (ii) interaction occurs only between the first and third components. Such an asymmetric interaction can potentially give rise to pairing superfluidity between the first and third components. Meanwhile, the asymmetric SOC modifies the momentum, energy, and population fraction of the first component. This manipulation is expected to have a significant impact on the possible pairing between the first and third components~\cite{zhou2014three}.
We apply mean-field theory to study the possible pairing. To proceed, we introduce the order parameter
\begin{equation} \Delta_{13}=\frac{U_{13}}{S}\sum_{\bm{k}}\left\langle c_{\bm{Q}-\bm{k},1}c_{\bm{k},3}\right\rangle.
\label{eq6}
\end{equation} 
It is the pair between the first and third components with a finite center-of-mass momentum $\bm{Q}$, i.e., FF pairing. With this FF order parameter, the interacting $H_\text{I}$ can be approximated as
\begin{align}
H_\text{I} = \sum_{\bm{k}}\left(\Delta^*_{13}c_{\bm{Q}-\bm{k},1}c_{\bm{k},3}+\Delta_{13}c_{\bm{k},3}^{\dagger}c_{\bm{Q}-\bm{k},1}^{\dagger}\right)-\frac{S\left|\Delta_{13}\right|^{2}}{U_{13}}. \notag 
\end{align}
We further construct the Nambu basis $\psi_{\bm{Q}}(\bm{k})=(
c_{\bm{k},1}, c_{\bm{k},2},  c_{\bm{k},3}, c_{\bm{Q}-\bm{k},1}^{\dagger}, c_{\bm{Q}-\bm{k},2}^{\dagger}, c_{\bm{Q}-\bm{k},3}^{\dagger} )^{T}$. With the mean-field approximation in $H_\text{I}$,  the total Hamiltonian in Eq.~(\ref{Hamiltonian}) can be expressed in the Nambu basis space as ~\cite{de2009three, zhou2014three, qin2015three},
\begin{align}
H=&\frac{1}{2}\sum_{\bm{k}}\psi_{\bm{Q}}^{\dagger}\left(\bm{k}\right)M_{\bm{Q}}(\bm{k})\psi_{\bm{Q}}\left(\bm{k}\right) \notag \\
&+\frac{1}{2}\sum_{\bm{k}}\left(3\varepsilon_{\bm{Q}-\bm{k}}-2 \mu_{12}-\mu_{3}\right)-\frac{S\left|\Delta_{13}\right|^{2}}{U_{13}}.
    \label{MeanFieldHamiltonian}
\end{align}
Here, the Bogoliubov–de-Gennes (BdG) matrix is
\begin{equation} 
M_{\bm{Q}}(\bm{k})=
\begin{pmatrix}
 H_0(\bm{k} ) & \Delta \\ \Delta^\dagger &    -H_0(\bm{Q}-\bm{k} )\end{pmatrix},
\label{BdGMatrix}
\end{equation}
with the matrix $\Delta$ being
\begin{equation} 
      \Delta=
\begin{pmatrix}
0 & 0& -\Delta_{13}\\
 0 & 0 & 0 \\
\Delta_{13} & 0 & 0 \\\end{pmatrix}.
\label{DeltaMatrix}
\end{equation}
From the Hamiltonian $H$ in Eq.~(\ref{MeanFieldHamiltonian}), we calculate the thermodynamic potential $\Omega =-k_\text{B}T \ln \text{Tr} e^{-H/(k_\text{B}T )}|_{T\rightarrow 0}$, here, $k_\text{B}$ is the Boltzmann constant and $T$ is the temperature. The calculated result becomes,
\begin{align}
\label{Thermodynamic}
    \Omega=&-\frac{1}{2}\sum_{\bm{k},n} E_{\bm{Q}}^{(n)}(\bm{k}) \Theta\left[ E_{\bm{Q}}^{(n)}(\bm{k}) \right]  \\    &+\frac{1}{2}\sum_{\bm{k}}\left(3\varepsilon_{\bm{Q}-\bm{k}}-2 \mu_{12}-\mu_{3}\right) 
    +\sum_{\bm{k}}\frac{\left|\Delta_{13}\right|^{2}}{2\varepsilon_{\bm{k}}+E_{b,13}}, \notag 
\end{align}
where $\Theta$ is the Heaviside step function, and $E_{\bm{Q}}^{(n)}(\bm{k})$ with $n=1-6$ are eigenvalues of the BdG matrix $M_{\bm{Q}}(\bm{k})$.

The thermodynamic potential plays a central role in this study. For given chemical potentials $\mu_{12}$ and $\mu_3$
, the ground state should correspond to the global minimum of the thermodynamic potential. We first locate this minimum and, from it, determine the pairing momentum  $\bm{Q}$ and the order parameter $\Delta_{13}$.  All results indicate that  $\bm{Q}$ is always along the $\hat{k}_x$ direction whenever it is nonzero:
\begin{equation}
\bm{Q}= Q_x \hat{k}_x,
\end{equation}
where $ Q_x$ denotes the magnitude of the pairing momentum along the $\hat{k}_x$ direction.
 This is reasonable because SOC plays an important role in the formation of FF pairs and is present only along the 
$\hat{k}_x$ direction. The asymmetric three-component system offers a fundamental opportunity to tune the population imbalance between the pairing components by varying the chemical potentials. In the following, we fix the spin-orbit-coupled chemical potential 
$\mu_{12}$ and vary $\mu_3$ to investigate FF pairing. We identify different types of pairing, which can be characterized by $ Q_x$ and $\Delta_{13}$.  For convenience in numerical calculations,  we  { adopt the units of momentum and energy as
${k}_{F3}$ and 	$E_{k_{F3}}=\hbar^2{k}_{F3}^2/2m$ , respectively, in order to make all quantities dimensionless.  Here, ${k}_{F3}$ is the Fermi momentum associated with the Fermi surface of the third component. 
}

\begin{figure*}[t]
 \centering
    \includegraphics[width=1 \textwidth]{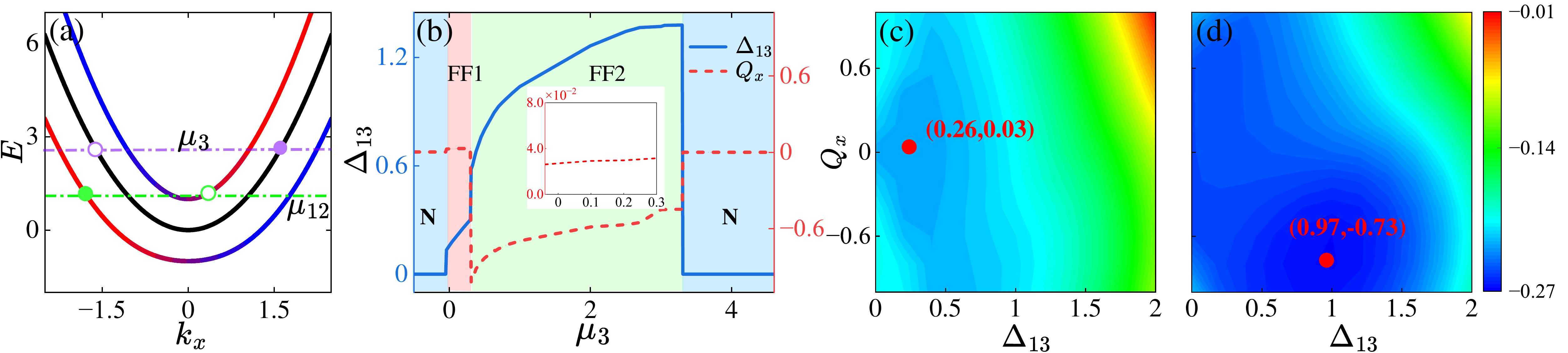}
  \caption{ FF pairs in an asymmetric three-component Fermi gas with weak SOC, $\alpha=1$, $h=1$, and $E_{b,13}=0.5$. (a) Single-particle dispersion of $H_0(\bm{k})$ as a function of $k_x$ for $k_y=0$. The isolated energy band of the third component is shown by the black line. The spin-orbit-coupled energy bands of the first and second components are shown by the colored lines. The red (blue)-colored parts emphasize the dominant occupation of the first (second) component. The green-dashed horizontal line represents $\mu_{12}=1.1$, which cuts through both the spin-orbit-coupled bands, generating an upper-band state (labeled by a green open circle) and a lower-band state (labeled by a green solid circle). The chemical potential $\mu_3$ cuts through the third-component band, resulting in two states (labeled by purple open and solid circles). (b) Phase diagram showing the evolution of $\Delta_{13}$ (blue solid line) and $Q_x$ (red-dashed line) of the ground states as functions of $\mu_3$. The diagram includes the normal phase (N, light blue region), the FF1 phase (light red region), and the FF2 phase (light green region). The inset shows a zoom-in of $Q_x$ in the FF1 phase. (c) and (d) show the thermodynamic potential $\Omega$ in the parameter space of $(\Delta_{13}, Q_x)$ for $\mu_3 = 0.1$ and $\mu_3 = 0.8$, respectively. The global minimum of the thermodynamic potential is labeled by a red circle.
  } 
   \label{Weak}
\end{figure*}

\section{FF pair for weak SOC}
\label{WeakSEC}

The SOC parameters include the spin-orbit coupling strength $\alpha$ and the Raman coupling strength $h$. Depending on whether $\alpha$  dominates or not, the system resides in either the weak or strong SOC regime, where the spin-orbit-coupled single-particle spectrum exhibits a dramatic difference.

We first study the weak SOC case. Figure~\ref{Weak}(a) shows the single-particle dispersion of $H_0(\bm{k})$ in Eq.~(\ref{SingleParticleH}). The black line represents the energy band of the third component, while colorful lines correspond to the spin-orbit-coupled energy bands of the first and second components. The color scale in these lines denotes the population fraction, with red (blue) indicating dominance of the first (second) component.  We fix the chemical potential $\mu_{12}=1.1$ (green-dashed line), which intersects two spin-orbit-coupled bands and yields two states dominated by the first component.  The state in the upper (lower) spin-orbit-coupled band is marked by the green open (solid) circle. Clearly, these two states have different Fermi surface sizes, with the lower-band state exhibiting a larger Fermi surface. Meanwhile, we choose the chemical potential 
 $\mu_3$ such that it produces two states (represented by purple solid and open circles) after intersecting the dispersion of the third component. These two states lie on the same Fermi surface. We anticipate that FF pairing occurs among these four labeled states.

Once the chemical potentials $\mu_{12}$ and $\mu_3$ are given, we calculate the thermodynamic potential $\Omega$ in Eq.~(\ref{Thermodynamic}) for various values of parameters $Q_x$ and $\Delta_{13}$. Two representative results in the parameter space of $(\Delta_{13},Q_x)$ for $\mu_3=0.1$ and $\mu_3=0.8$ are shown in 
Figs.~\ref{Weak}(c) and \ref{Weak}(d), respectively. From the calculated $\Omega$, we identify the locations of its minima, which are marked by red circles. These minima correspond to the ground state of the system.  In each minimum, if $\Delta_{13}=0$, the ground state is in the normal phase (N); if $\Delta_{13} \ne 0$ and $Q_x \ne 0$, the ground state is in the FF pairing superfluid phase.  We find that the FF pairs in Figs.~\ref{Weak}(c) and \ref{Weak}(d) exhibit different behavior,and thus we designate them as FF1 and FF2 phases, respectively. Interestingly, for this asymmetric three-component system, there is no conventional superfluid phase characterized by $\Delta_{13} \ne 0$ and $Q_x = 0$~\cite{zhou2014three}. 

We fix the chemical potential $\mu_{12}=1.1$ and vary $\mu_3$, then repeat the above calculation of the thermodynamic potential to determine the values of $\Delta_{13}$ and $Q_x$ for the ground states. These values are plotted as functions of $\mu_3$ in Fig.~\ref{Weak}(b).

(1) When $\mu_3$ is negative, the system is in the normal phase (N). In this regime, $\mu_3$ does not intersect the energy band of the third component; consequently, the finite binding energy $E_{b,13}$ is insufficient to bind the first and third components together into pairs.

(2) As $\mu_3$ increases and eventually intersects the band of the third component, the system enters the FF1 phase (pink-shaded region). This phase is characterized by a nonzero 
$\Delta_{13}$  and a very small positive  $Q_x$ [see the inset in Fig.~\ref{Weak}(b)]. The emergence of such a small positive $Q_x$ provides an important clue to the pairing mechanism. In the FF1 region, 
$\mu_3$ cuts the energy band of the third component, generating a small Fermi surface. Meanwhile, the Fermi surface of the upper-band state [indicated by the green open circle in Fig.~\ref{Weak}(a)] is also very small, in contrast to the larger Fermi surface of the lower-band state [green solid circle in Fig.~\ref{Weak}(a)]. The comparable sizes of the Fermi surfaces of the upper-band state and the third-component state facilitate pairing between them and naturally lead to a very small pairing momentum. The order parameter $\Delta_{13}$ increases with $\mu_3$, as a higher $\mu_3$
enhances the density of the third component.

(3) As $\mu_3$ increases further, the system enters the FF2 phase [light-green-shaded region in Fig.~\ref{Weak}(b)]. The phase transition between the FF1 and FF2 phases is first order. In the FF2 phase, the pairing momentum 
$Q_x$ is no longer small, and both $Q_x$ and $\Delta_{13}$ increase with $\mu_3$. In this regime, the Fermi surface of the lower-band state becomes comparable in size to that of the third component, facilitating pairing between them [see the two solid circles in Fig.~\ref{Weak}(a)]. Moreover, the Fermi surface of the third component lies inside that of the lower-band state, which gives rise to a negative pairing momentum $Q_x$. 
As $\mu_3$ continues to increase, the Fermi surface of the third component expands, causing the pairing momentum to become less negative.

{ (4) For sufficient large $\mu_3$, the FF2 phase disappears and the system becomes in the normal phase again, generating a first-order transition transition between the FF2 and normal phases as shown in Fig.~\ref{Weak}(b). The finite binding energy can not bind the first and third components together to form pairs for sufficient large $\mu_3$ due to the large mismatch of Fermi surfaces of the first and third components.  }

We show that the FF1 (FF2) phase corresponds to the formation of pairs between the third component and the upper- (lower-) band spin-orbit-coupled state. By tuning 
$\mu_3$, a { first-order }phase transition between these two phases can be induced.  {
As demonstrated above, the first-order transition arises from the population of the first component swapping from the upper spin-orbit-coupled band to the lower one. The thermodynamic potential $\Omega$ exhibits two minima in the parameter space: one is a local minimum, and the other is the global minimum. The FF pairing is associated with the global minimum. The phase transition corresponds to the swapping between the local and global minima.	} We emphasize that the order parameter $\Delta_{13}$ in the FF2 phase is always larger than that in the FF1 phase. This is because the Fermi surface of the lower-band state is larger than that of the upper-band state, resulting in a higher density in the lower-band state.  {Finally, we comment that the FF superfluids can exist in a broad range of $\mu_3$. The reason is that the first-component pairing states originate only from the left side in the lower spin-orbit-coupled band. It is the spin-orbit coupling that guarantee this asymmetry. If the first-component pairing states were also available from the right side in the lower band, the effective system would be similar to a spin-imbalanced Fermi gas, where the FF pairs can exist only in an extremely narrow parameter region.  }

\begin{figure}[b]
 	\centering
 	\includegraphics[width=0.35 \textwidth]{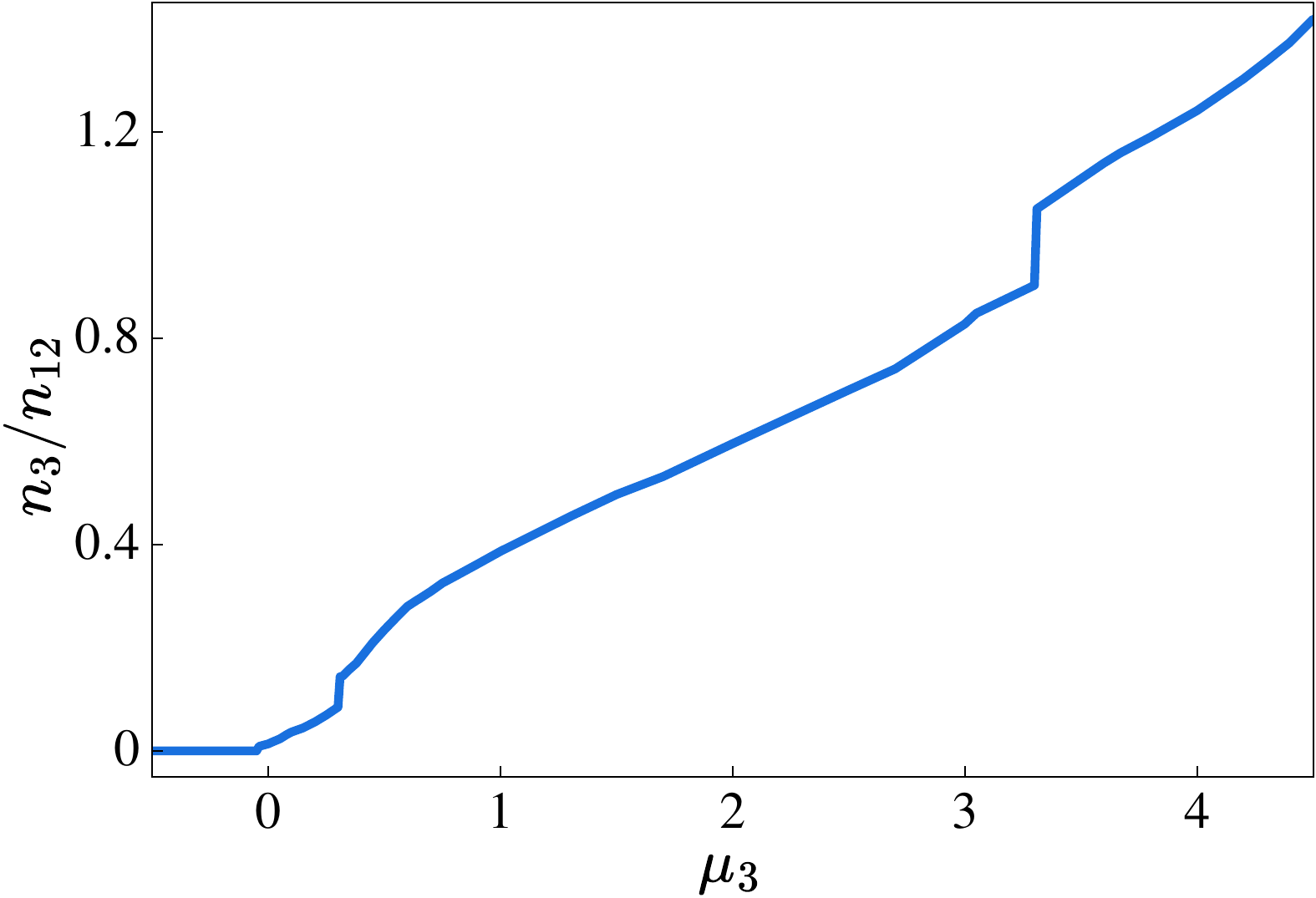}
 	\caption{The  concentration $n_3/n_{12}$ in an asymmetric three-component Fermi gas with weak SOC. All parameters are same as Fig.~\ref{Weak}(b).} 
 	\label{Concentration_wake}
 \end{figure}

{ In real experiments, the tunability of  the chemical potential is realized by controlling the density.  This stimulates us to study the relation between the density and chemical potential. The densities are calculated from the thermodynamic potential as
\begin{align}
n_3=-\frac{\partial \Omega}{\partial \mu_3}, \ \ \ 
n_{12}=-\frac{\partial \Omega}{\partial \mu_{12}}.
\end{align}
Here, $n_3$ and $n_{12}$ denote the densities of the third and first components, respectively. The concentration is defined as 
$n_{3}/n_{12}$.  Using the same parameters as in Fig.~\ref{Weak}(b), we calculate the concentration, and the results are shown in Fig.~\ref{Concentration_wake}. Notable features include sudden jumps in the concentration at the transitions between different phases. Additionally, in the normal phase with a small $\mu_3$ , the third component contains no atoms. These discontinuities in the concentration can be used for experimental detection of phase transitions.

}

\section{FF pair for strong SOC}
\label{StrongSEC}

Reference~\cite{zhou2014three} studied the same weak SOC case and reported results similar to ours. In the following, we investigate the strong SOC case, which goes beyond the study in Ref.~\cite{zhou2014three}, and identify the existence of a new type of FF pair. The strong SOC gives rise to a double-well structure in the lower spin-orbit-coupled band, as shown in Fig.~\ref{Stronge}(a1). We find that the choice of $\mu_{12}$ has a significant effect on FF pairs.

\subsection{ $\mu_{12}$ cuts through both spin-orbit-coupled bands}

We first study the case where $\mu_{12}$ cuts through both spin-orbit-coupled bands. As indicated by the green solid and open circles in Fig.~\ref{Stronge}(a1), this choice of $\mu_{12}$ yields two available states that are dominated by the first component [see the red-colored line]. In Fig.~\ref{Stronge}(a2), $Q_x$ and $\Delta_{13}$ of the ground states are shown as functions of $\mu_3$. The phase diagram is similar to that in Fig.~\ref{Weak}(b): The pink region corresponds to the FF1 phase, in which FF pairing occurs between the upper-band state (green open circle) and the third component; the FF2 phase (light-green region), on the other hand, corresponds to FF pairing between the lower-band state (green solid circle) and the third component.

\subsection{ $\mu_{12}$ lies within the spin-orbit-coupled gap }

\begin{figure}[t]
 \centering
\includegraphics[width=0.48 \textwidth]{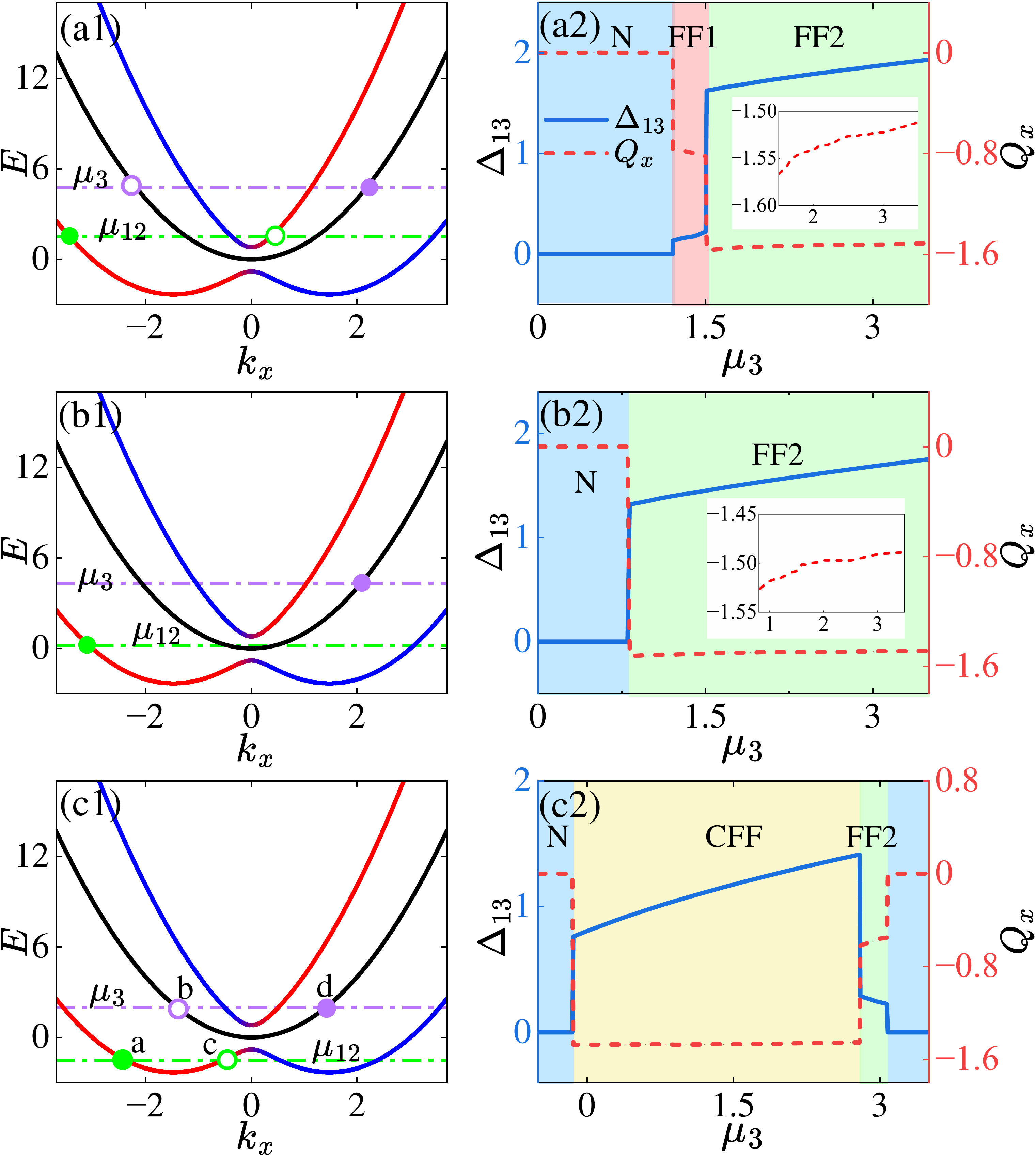}
\caption{
FF pairs in an asymmetric three-component Fermi gas with strong SOC, $\alpha=3$, $h = 0.8$, and $E_{b,13} = 0.5$. The left panels show the single-particle energy bands, and the right panels show the corresponding phase diagrams. (a1) and (a2) $\mu_{12} = 1.5$ cuts through both spin-orbit-coupled bands. (b1) and (b2) $\mu_{12} = 0.2$ lies within the spin-orbit coupling gap. The insets in (a2) and (b2) provide zoomed-in views of $Q_x$ to illustrate its variation. (c1) and (c2) $\mu_{12} = -1.5$ cuts through the double-well structure within the lower spin-orbit-coupled band. The new composite FF phase, labeled ``CFF", is shown in the light yellow region. } 
    \label{Stronge}
\end{figure}

If $\mu_{12}$ is chosen to lie inside the spin-orbit-coupled gap, as shown in Fig.~\ref{Stronge}(b1), only one first-component-dominated state remains in the lower band (green solid circle). This state can bind with the third-component state (purple solid circle) to form an FF2 pair. Since no upper-band state is cut by $\mu_{12}$, the FF1 pair cannot exist. The phase diagram presented in Fig.~\ref{Stronge}(b2) shows that increasing $\mu_3$ leads to a first-order phase transition directly from the normal phase to the FF2 phase, without passing through the FF1 phase.

\subsection{ $\mu_{12}$ cuts through the double-well structure in the lower spin-orbit-coupled band}

The results for the two choices of $\mu_{12}$ discussed above are very similar to those for the weak SOC case presented in the previous section. However, when $\mu_{12}$ is chosen to intersect the double-well structure in the lower spin-orbit-coupled band, no analogue exists in the weak SOC regime. This parameter configuration thus represents a unique feature of the strong SOC case.

Fig.~\ref{Stronge}(c1) shows that when $\mu_{12}$ intersects the double-well structure, two first-component-dominated states are generated, labeled by the green solid circle ``a" and the green open circle ``c". Meanwhile, $\mu_3$ gives rise to two possible states, marked by the purple open circle ``b" and the purple solid circle ``d". The ``a" and ``d" states can form a pair with a finite net pairing momentum $q_1$, as these two states have opposite group velocities. For the same reason, the ``b" and ``c" states can form a pair with a finite net pairing momentum $q_2$. The { composition} of these two distinct pairs gives rise to a new FF state with a pairing momentum $Q_x=(q_1+q_2)/2$ in order to minimize the energy. We name this new state the CFF state. The phase diagram shown in Fig.~\ref{Stronge}(c2) indicates that the CFF phase can exist over a wide range of $\mu_3$. A distinct feature of the CFF phase is that the pairing momentum $Q_x$ remains constant ($Q_x=-1.47$ for the parameters in Fig.~\ref{Stronge}(c2)) throughout the entire region where it exists. We have checked other values of the chemical potential $\mu_{12}$ that cut through the double-well structure and found that the CFF phase consistently exhibits $Q_x=-1.47$. In the CFF phase, an increase in $\mu_3$ causes the momenta $k_x$ of the ``b" and ``d" states to expand symmetrically in opposite directions. Consequently, $q_1$ in the pair formed by the ``a" and ``d" states increases by an amount exactly equal to the decrease in $q_2$ in the pair formed by the ``b" and ``c" states. As a result, the average pairing momentum $Q_x=(q_1+q_2)/2$ remains invariant as $\mu_3$ varies.
{ If we consider a negative strong SOC strength ($\alpha=-3$), the pairing momentum $Q_x$ in the CFF phase shifts to a positive value $Q_x=1.47$. }

\begin{figure}[t]
	\centering
	\includegraphics[width=0.4 \textwidth]{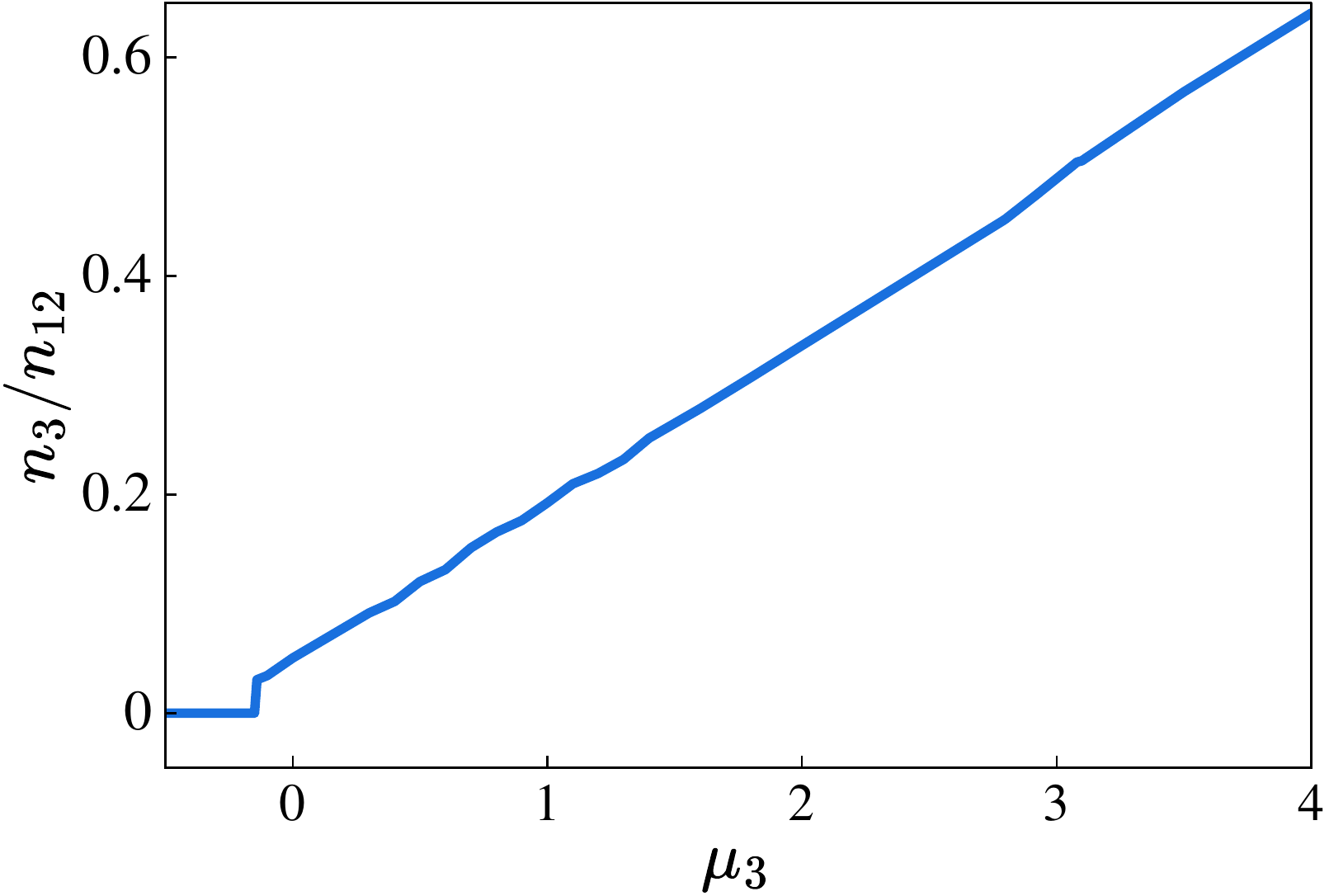}
	\caption{The concentration $n_3/n_{12}$ in an asymmetric three-component Fermi gas with strong SOC. All parameters are same as Fig.~\ref{Stronge}(c2).} 
	\label{Concentration_strong}
\end{figure}

{
The CFF state is novel due to the existence of the momentum-space double-well structure in the lower spin-orbit-coupled band. In the following, we provide a further simple picture to explain why the state exists in the form of ``composition" as described above.  From Fig.~\ref{Stronge}(c1), it appears that the states ``a" and ``c"
arise from a parabolic-like dispersion 
$(k_x+\beta)^2/(2m_\text{eff})$ with $m_\text{eff}$, where  $m_\text{eff}$
is an effective mass that mimics the curvature of the dispersion (shown by the red-colored line in the lower spin-orbit-coupled band), and  $\beta$ 
describes the center of the parabola ($\beta=1.47$ in Fig.~\ref{Stronge}(c1)). We further assume $m_\text{eff}\approx 1$.
Meanwhile, states  ``b" and  ``d" originate from the dispersion of the third component  $k_x^2/2$.  Consequently, these four states can be effectively described by the single-particle Hamiltonian
\begin{equation}
H_\text{eff}(\bm{k}, \bm{\beta})=\begin{pmatrix}\frac{(k_x+\beta)^2}{2} +\frac{k_y^2}{2} -\mu_{12} & 0 \\ 0 & \frac{k_x^2}{2}+\frac{k_y^2}{2}-\mu_3
\end{pmatrix}.
\end{equation}
Here, $\bm{\beta}=(\beta,0)$. 
Considering the interactions, the total effective system becomes
\begin{align}
 H'&=\sum_{\bm{k}} \Psi^\dagger(\bm{k}) H_\text{eff}(\bm{k},\bm{\beta}) \Psi(\bm{k}) \notag \\
 &+ \sum_{\bm{q},\bm{k},\bm{k'}}\left(C_{\bm{k},3}^{\dagger}C_{\bm{q}-\bm{k},1}^{\dagger}C_{\bm{q}-\bm{k'},1}C_{\bm{k'},3}\right),
 \label{TotalHamitlonian}
\end{align}
with  $\Psi(k)=(
C_{\bm{k},1},  C_{\bm{k},3})^T$ and $
C_{\bm{k},1}$ ($C_{\bm{k},3}$) describing the states ``a" and ``c" (``b" and ``d"). The mean-field order parameter is introduced as $ \Delta_{13}(\bm{Q})=\sum_{\bm{k}}\left\langle C_{\bm{Q}-\bm{k},1}C_{\bm{k},3}\right\rangle
$, which still features FF paring if momentum $\bm{Q}\neq 0$. Under the mean-field approximation and in the Nambu basis $\psi_{\bm{Q}}(\bm{k})=(
C_{\bm{k},1}, C_{\bm{k},3}, C_{\bm{Q}-\bm{k},1}^{\dagger},  C_{\bm{Q}-\bm{k},3}^{\dagger} )^{T}$, the Hamiltonian turns to be
\begin{align}
H'& =\frac{1}{2}\sum_{\bm{k}}\psi_{\bm{Q}}^{\dagger}\left(\bm{k}\right) \begin{pmatrix}
H_\text{eff}(\bm{k},\bm{\beta}) & \Delta(\bm{Q} ) \\ \Delta^\dagger(\bm{Q} ) &    -H_\text{eff}(\bm{Q}-\bm{k},\bm{\beta}) \end{pmatrix}\psi_{\bm{Q}}\left(\bm{k}\right) \notag \\
&+\frac{1}{2}\sum_{\bm{k}}
\left(\frac{(\bm{Q}-\bm{k}+\bm{\beta} )^2 +  (\bm{Q}-\bm{k})^2 }{2}
- \mu_{12}-\mu_{3}\right) \notag \\
&-|\Delta_{13}(\bm{Q})|^2.
\end{align}
Here the matrix 
\begin{equation}
\Delta(\bm{Q} )= \begin{pmatrix}
0&-\Delta_{13}(\bm{Q})\\ \Delta_{13}(\bm{Q})&0
\end{pmatrix}.
\end{equation}
We perform a unitary transformation: $\psi_{\bm{Q}}(\bm{k})\rightarrow \psi_{\bm{Q}}^{\bm{\beta}}(\bm{k})=(
e^{i\beta x}C_{\bm{k},1}, C_{\bm{k},3}, e^{-i\beta x}C_{\bm{Q}-\bm{k},1}^{\dagger},  C_{\bm{Q}-\bm{k},3}^{\dagger} )^{T}$.  Then the mean-field system becomes
\begin{align}
&H'= \notag \\ &\frac{1}{2}\sum_{\bm{k}}\psi_{\bm{Q}}^{\bm{\beta}\dagger}(\bm{k}) \begin{pmatrix}
H_\text{eff}(\bm{k},0) & \Delta_\text{eff}(\bm{Q} ) \\ \Delta_\text{eff}^\dagger(\bm{Q} ) &    -H_\text{eff}(\bm{Q}-\bm{k},0) \end{pmatrix}\psi_{\bm{Q}}^{\bm{\beta}}(\bm{k}) \notag \\
&+\frac{1}{2}\sum_{\bm{k}}
\left(\frac{(\bm{Q}-\bm{k}+\bm{\beta} )^2 +  (\bm{Q}-\bm{k})^2 }{2}
- \mu_{12}-\mu_{3}\right) \notag \\
&-|\Delta_{13}(\bm{Q})|^2,
\label{NeweffectiveHamiltonian}
\end{align}
with $\Delta_\text{eff}(\bm{Q} )=e^{i\beta x} \Delta(\bm{Q} )$. It is noticed that the system in Eq.~(\ref{NeweffectiveHamiltonian}) describes a spin-imbalanced Fermi gas where the imbalance is controlled by $\mu_{12}-\mu_3$. It has been widely studied in literature~\cite{radzihovsky2010imbalanced,sheehy2015fulde,kinnunen2018fulde,he2006loff}. The outstanding results of the system include that the ground state is the normal superfluid, i.e.,  $\Delta_\text{eff}(\bm{Q} )\neq 0$ but with $\bm{Q}=0$ , in wide parameter regions, and the FF superfluid $\Delta_\text{eff}(\bm{Q} )\neq 0$ with $\bm{Q} \neq 0$ only exists in a particular narrow region of $\mu_{12}-\mu_3 $ and furthermore the pairing momentum $\bm{Q}$ remains small values . Now transforming the superfluid phases (including both the normal superfluid and the FF superfluid) of the system in Eq.~(\ref{NeweffectiveHamiltonian}) back into the space of  bare states shown by ``a", ``c", ``b" and ``d", one immediately realizes 
\begin{equation}
\Delta(\bm{Q} )=e^{-i\beta x} \Delta_\text{eff}(\bm{Q} ).
\end{equation}
Since other $\Delta_\text{eff}(\bm{Q} )\neq 0$ with $\bm{Q}=0$ or $\Delta_\text{eff}(\bm{Q} )\neq 0$ with a small finite $\bm{Q}$, we naturally conclude that the ground state of the effective system in Eq.~(\ref{TotalHamitlonian}) is in the FF superfluid with the pairing momentum $Q_x\approx -\beta$ in very wide parameter regions. Such the FF superfluid is the CFF superfluid revealed in Fig.~\ref{Stronge}(c2)) and the pairing momentum $Q_x\approx -\beta$ is invariant if $\mu_{12}-\mu_3$ varies. }

{ From the above analysis, we know that it is the relative center displacement between the two parabolic dispersions that induces the pairing momentum for the CFF states. Such asymmetry between the two dispersions is controlled by the spin-orbit coupling. }

Furthermore, Fig.~\ref{Stronge}(c2) shows a first-order phase transition from the CFF phase to the FF2 phase as $\mu_3$ increases. In the FF2 phase, the pairing momentum $q_1$ in the FF pair formed by the ``b" and ``c" states becomes largely negative, rendering this FF pair energetically unfavorable. Consequently, the composite FF pair collapses, leaving only the FF pair formed by the ``a" and ``d" states. Compared to the CFF phase, the FF2 phase exists only in a narrow region of $\mu_3$. As $\mu_3$ increases further, it is followed by the normal phase. { By calculating the concentration, we find that the discontinuities of the concentration in the transitions between the CFF and FF2 phases and between the FF2 and the normal phases are not very obvious [see Fig.~\ref{Concentration_strong}].  }

\begin{figure}[t]
	\centering
	\includegraphics[width=0.4 \textwidth]{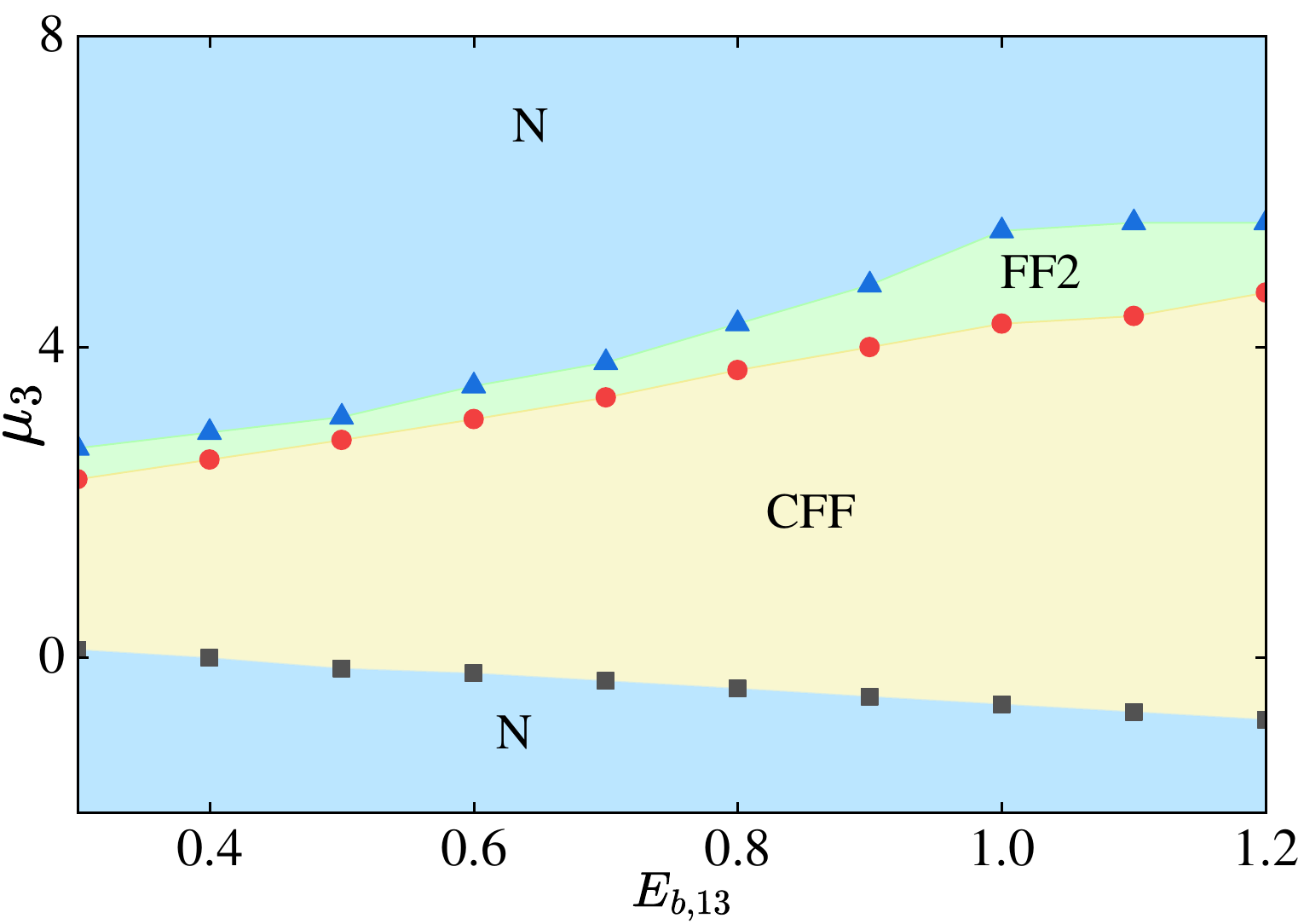}
	\caption{Phase diagram in the ($E_{b,13},\mu_3$) plane in an asymmetric three-component Fermi gas with strong SOC. The parameters are the same as those in Fig.~\ref{Stronge} with $\mu_{12}=-1.5$.} 
	\label{CFFbinding}
\end{figure}

We emphasize that the CFF phase can exist over a very broad range of the binding energy $E_{b,13}$. A phase diagram in the ($E_{b,13}, \mu_{3}$) plane is presented in Fig.~\ref{CFFbinding}. Interestingly, the region of the CFF phase along the $\mu_{3}$ dimension expands as $E_{b,13}$ increases. The binding energy helps inhibit the collapse of the composite FF pairs. Meanwhile, the FF2 phase continues to exist throughout. { Finally, we would like to point out that the pairing momentum in the FF1 and FF2 phases depends on the binding energy, which is consistent with existing studies~\cite{xu2015topological,qu2013topological,xu2014anisotropic,zheng2013route,zheng2014fflo,zhou2014three}. However, for the CFF phase, the pairing momentum $Q_x$ only associates with the spin-orbit coupling strength (resulting in $\beta$ in above) and does not depends on the binding energy. Therefore, the composite states in the CFF phase represented by the light yellow region in  Fig.~\ref{CFFbinding} have the same pairing-momentum value. }

\begin{figure}[t]
	\centering
	\includegraphics[width=0.4 \textwidth]{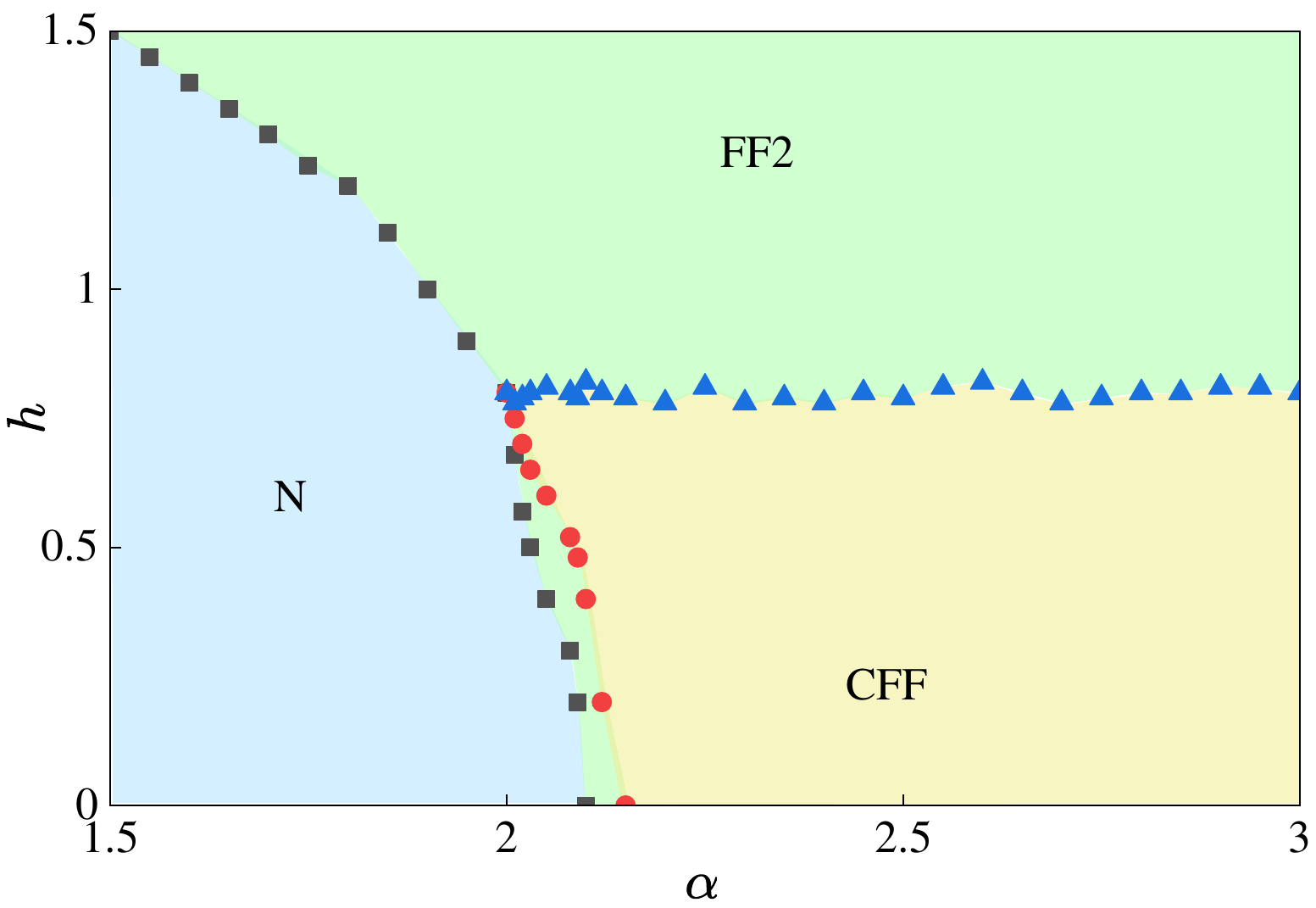}
	\caption{Phase diagram in the ($\alpha$,$h$) plane in an asymmetric three-component Fermi gas. The parameters are chosen as $\mu_{12}=-1$, $\mu_3=1.5$, and  $E_{b,13}=0.5$.} 
	\label{H-alpha}
\end{figure}

\section{FF pair in parameter space of SOC}

{
In previous sections, we have revealed the existence of FF superfluids in both weak and strong SOC regimes. In spin-orbit-coupled atomic experiments, the SOC strength $\alpha$ can be tuned by adjusting the incident angle between Raman lasers, and the two-photon Rabi coupling  $h$ can be experimentally modified by changing the intensities of the Raman lasers~\cite{wang2012spin,cheuk2012spin,lin2011spin}. The FF superfluids depend strongly on the SOC parameters. Therefore, it is interesting to study the phase diagram in the SOC parameter space ($h,\alpha$).

 Fig.~\ref{H-alpha} shows such a typical phase diagram, where $\mu_{12}$ only cuts the lower spin-orbit-coupled band,  so that the FF1 phase does not appear. The chemical potential $\mu_{3}$ and the binding energy $E_{b,13}$ are fixed. With these fixed parameters, the normal phase is located in the region of  small $\alpha$ [see the light-blue region in Fig.~\ref{H-alpha}].  For large $alpha$, the ground state becomes the CFF phase if $h$ is small. Increasing $h$ drives a transition from the CFF phase to the FF2 phase. These findings are consistent with the results shown in Figs.~\ref{Weak} and ~\ref{Stronge}. Between the normal and CFF phases there exists a narrow parameter region where the ground state is in the FF2 phase [see the light-green region surrounded by squares and circles]. In this parameter region, the lower spin-orbit-coupled energy band becomes nearly flat around $k=0$, because the double wells are very close to each other in momentum space.  In this case,  The chemical potential $\mu_{12}$ can only generate a single pairing state, which couples with the third component to form the FF2 state. 
}

{ In a very recent experiment, a three-component Fermi gas has been realized experimentally~\cite{schumacher2026observation}. Given the fully experimental tunability of spin-orbit coupling in the aforementioned experiments~\cite{wang2012spin,cheuk2012spin,lin2011spin}, we expect that the asymmetric three-component Fermi gas can be experimentally implemented. The observation of the pairing momentum will distinguish which FF superfluid the ground state belongs to, as the FF1 and FF2 phases feature different magnitudes of the pairing momentum, while it remains invariant in the CFF phase.  }

\section{Conclusions}
\label{Conclusion}

An asymmetric three-component Fermi gas provides an important platform for investigating exotic FF superfluids by combining SOC with population imbalance (via chemical potential mismatch). We systematically study the ground-state phase diagram of such an asymmetric three-component { two dimensional} Fermi gas by finding the global minima of the thermodynamic potential. Strong SOC gives rise to a double-well structure in momentum space within the lower spin-orbit-coupled band. By choosing the chemical potential to intersect this double-well structure, we reveal a new composite FF superfluid. This composite FF superfluid exists over a wide range of parameters, and its pairing momentum remains invariant under changes in the chemical potentials.

{The existence of the composite FF superfluid is revealed based on the assumption of the FF order-parameter ansatz. It would be interesting to examine whether it remains the ground state when the LO order parameter is assumed instead. On the other hand, thanks to the versatile tunability of degenerate Fermi gases, the BCS-BEC crossover can be achieved experimentally. How the composite FF superfluid, which exists in the BCS regime, evolves when the interaction is tuned to the BEC side presents a potential research direction. }

\begin{acknowledgments}
This work is supported by the National Natural Science Foundation of China (NSFC) under Grants No.12374247 and No.11974235, by the Shanghai Municipal Science and Technology Major Project (Grant No. 2019SHZDZX01-ZX04), and the Shanghai Science and Technology Innovation Action Plan (Grant No. 24LZ1400800), as well as by the State Key Laboratory of Micro-nano Engineering Science (Grant No. MES202605).
\end{acknowledgments}


\begin{thebibliography}{60}
\bibliographystyle{unsrt}

\bibitem{RevModPhys.80.885} I. Bloch, J. Dalibard, and W. Zwerger, Many-body physics with ultracold gases, \href{https://doi.org/10.1103/RevModPhys.80.885}{Rev. Mod. Phys. {\bf 80}, 885 (2008)}.

\bibitem{schrieffer2018theory} J. R. Schrieffer, \textit{Theory of superconductivity} (CRC press, 1999).


\bibitem{giorgini2008theory} S. Giorgini, L. P. Pitaevskii, and S. Stringari, Theory of ultracold atomic Fermi gases, \href{https://doi.org/10.1103/RevModPhys.80.1215}{Rev. Mod. Phys. {\bf 80}, 1215 (2008)}.


\bibitem{schumacher2026observation} Schumacher, Grant L and M{\"a}kinen, Jere T and Ji, Yunpeng and T. Assump{\c{c}}{\~a}o, Gabriel G and Chen, Jianyi and Huang, Songtao and Vivanco, Franklin J and Navon, Nir, Observation of anomalous decay of a polarized three-component Fermi gas, \href{https://www.nature.com/articles/s41467-025-65183-3}{Nat. Commun. {\bf 17}, 174 (2026)}.



\bibitem{strinati2018bcs} G. C. Strinati, P. Pieri, G. R{\"o}pke, P. Schuck, and M. Urban, The BCS-BEC crossover: From ultra-cold Fermi gases to nuclear systems, \href{https://www.sciencedirect.com/science/article/pii/S0370157318300267}{Phys. Rep. {\bf 738}, 1 (2018)}.

\bibitem{randeria1990superconductivity} M. Randeria, J.-M. Duan, and L.-Y. Shieh, Superconductivity in a two-dimensional Fermi gas: Evolution from Cooper pairing to Bose condensation, \href{https://doi.org/10.1103/PhysRevB.41.327}{Phys. Rev. B {\bf 41}, 327 (1990)}.

\bibitem{marini1998evolution} M. Marini, F. Pistolesi, and G. C. Strinati, Evolution from BCS superconductivity to Bose condensation: analytic results for the crossover in three dimensions, \href{https://link.springer.com/article/10.1007/s100510050165}{Eur. Phys. J. B {\bf 1}, 151 (1998)}.

\bibitem{regal2004observation} C. A. Regal, M. Greiner, and D. S. Jin, Observation of Resonance Condensation of Fermionic Atom Pairs, \href{https://doi.org/10.1103/PhysRevLett.92.040403}{Phys. Rev. Lett. {\bf 92}, 040403 (2004)}.

\bibitem{zwierlein2004condensation} M. W. Zwierlein, C. A. Stan, C. H. Schunck, S. M. F. Raupach, A. J. Kerman, and W. Ketterle, Condensation of Pairs of Fermionic Atoms near a Feshbach Resonance, \href{https://doi.org/10.1103/PhysRevLett.92.120403}{Phys. Rev. Lett. {\bf 92}, 120403 (2004)}.

\bibitem{fulde1964superconductivity} P. Fulde and R. A. Ferrell, Superconductivity in a Strong Spin-Exchange Field, \href{https://doi.org/10.1103/PhysRev.135.A550}{Phys. Rev. {\bf 135}, A550 (1964)}.

\bibitem{larkin1965nonuniform} A. I. Larkin and Y. N. Ovchinnikov, Nonuniform state of superconductors, \href{https://www.semanticscholar.org/paper/Nonuniform-state-of-superconductors-Larkin-Ovchinnikov/09674bf67a436ef5797d48d46c9271714eb482f1}{Sov. Phys. JETP {\bf 20}, 762 (1965)}.

\bibitem{casalbuoni2004inhomogeneous} R. Casalbuoni and G. Nardulli, Inhomogeneous superconductivity in condensed matter and QCD, \href{https://doi.org/10.1103/RevModPhys.76.263}{Rev. Mod. Phys. {\bf 76}, 263 (2004)}.

\bibitem{radzihovsky2010imbalanced} L. Radzihovsky and D. E. Sheehy, Imbalanced Feshbach-resonant Fermi gases, \href{https://iopscience.iop.org/article/10.1088/0034-4885/73/7/076501}{Rep. Prog. Phys. {\bf 73}, 076501 (2010)}.

\bibitem{yi2015pairing} W. Yi, W. Zhang, and X. Cui, Pairing superfluidity in spin-orbit coupled ultracold Fermi gases, \href{https://link.springer.com/article/10.1007/s11433-014-5609-8}{Sci. China Phys. Mech. Astron. {\bf 58}, 1 (2015)}.

\bibitem{xu2015topological} Y. Xu and C. Zhang, Topological Fulde-Ferrell Superfluids of a Spin-Orbit Coupled Fermi Gas, \href{https://doi.org/10.1142/S0217979215300017}{Int. J. Mod. Phys. B {\bf 29}, 1530001 (2015)}.

\bibitem{paananen2006pairing} T. Paananen, J.-P. Martikainen, and P. T\"orm\"a, Pairing in a three-component Fermi gas, \href{https://doi.org/10.1103/PhysRevA.73.053606}{Phys. Rev. A {\bf 73}, 053606 (2006)}.

\bibitem{de2009three} T. N. De Silva, Three-component fermion pairing in two dimensions, \href{https://doi.org/10.1103/PhysRevA.80.013620}{Phys. Rev. A {\bf 80}, 013620 (2009)}.

\bibitem{ozawa2010population} T. Ozawa and G. Baym, Population imbalance and pairing in the BCS-BEC crossover of three-component ultracold fermions, \href{https://doi.org/10.1103/PhysRevA.82.063615}{Phys. Rev. A {\bf 82}, 063615 (2010)}.

\bibitem{sheehy2006bec} D. E. Sheehy and L. Radzihovsky, BEC-BCS Crossover in “Magnetized” Feshbach-Resonantly Paired Superfluids, \href{https://doi.org/10.1103/PhysRevLett.96.060401}{Phys. Rev. Lett. {\bf 96}, 060401 (2006)}.

\bibitem{zhang2005pairing} J. Zhang and H. Zhai, Pairing between atoms and molecules in a boson-fermion resonant mixture, \href{https://doi.org/10.1103/PhysRevA.72.041602}{Phys. Rev. A {\bf 72}, 041602 (2005)}.

\bibitem{tuzemen2025hierarchy} T{\"u}zemen, Bu{\u{g}}ra and Sowi{\'n}ski, Tomasz, Hierarchy of pairing in an imbalanced three-component one-dimensional Fermi gas, \href{https://journals.aps.org/pra/abstract/10.1103/yf54-cg95}{Phys. Rev. A {\bf 112}, 023322 (2025)}.




\bibitem{liao2010spin} Y.-a. Liao, A. S. C. Rittner, T. Paprotta, W. Li, G. B. Partridge, R. G. Hulet, S. K. Baur, and E. J. Mueller, Spin-imbalance in a one-dimensional Fermi gas, \href{https://www.nature.com/articles/nature09393}{Nature {\bf 467}, 567 (2010)}.

\bibitem{parish2007finite} M. M. Parish, F. M. Marchetti, A. Lamacraft, and B. D. Simons, Finite-temperature phase diagram of a polarized Fermi condensate, \href{https://www.nature.com/articles/nphys520}{Nat. Phys. {\bf 3}, 124 (2007)}.

\bibitem{zwierlein2006fermionic} M. W. Zwierlein, A. Schirotzek, C. H. Schunck, and W. Ketterle, Fermionic superfluidity with imbalanced spin populations, \href{https://www.science.org/doi/abs/10.1126/science.1122318}{Science {\bf 311}, 492 (2006)}.

\bibitem{partridge2006pairing} G. B. Partridge, W. Li, R. I. Kamar, Y.-a. Liao, and R. G. Hulet, Pairing and phase separation in a polarized Fermi gas, \href{https://www.science.org/doi/abs/10.1126/science.1122876}{Science {\bf 311}, 503 (2006)}.



\bibitem{wang2012spin} P. Wang, Z.-Q. Yu, Z. Fu, J. Miao, L. Huang, S. Chai, H. Zhai, and J. Zhang, Spin-Orbit Coupled Degenerate Fermi Gases, \href{https://doi.org/10.1103/PhysRevLett.109.095301}{Phys. Rev. Lett. {\bf 109}, 095301 (2012)}.

\bibitem{cheuk2012spin} L. W. Cheuk, A. T. Sommer, Z. Hadzibabic, T. Yefsah, W. S. Bakr, and M. W. Zwierlein, Spin-Injection Spectroscopy of a Spin-Orbit Coupled Fermi Gas, \href{https://doi.org/10.1103/PhysRevLett.109.095302}{Phys. Rev. Lett. {\bf 109}, 095302 (2012)}.

\bibitem{lin2011spin} Y.-J. Lin, K. Jim{\'e}nez-Garc{\'\i}a, and I. B. Spielman, Spin-orbit-coupled Bose-Einstein condensates, \href{https://www.nature.com/articles/nature09887}{Nature {\bf 471}, 83 (2011)}.

\bibitem{zhang2013topological} W. Zhang and W. Yi, Topological Fulde-Ferrell-Larkin-Ovchinnikov states in spin-orbit-coupled Fermi gases, \href{https://www.nature.com/articles/ncomms3711}{Nat. Commun. {\bf 4}, 2711 (2013)}.

\bibitem{qu2013topological} C. Qu, Z. Zheng, M. Gong, Y. Xu, L. Mao, X. Zou, G. Guo, and C. Zhang, Topological superfluids with finite-momentum pairing and Majorana fermions, \href{https://www.nature.com/articles/ncomms3710}{Nat. Commun. {\bf 4}, 2710 (2013)}.

\bibitem{zheng2013route} Z. Zheng, M. Gong, X. Zou, C. Zhang, and G. Guo, Route to observable Fulde-Ferrell-Larkin-Ovchinnikov phases in three-dimensional spin-orbit-coupled degenerate Fermi gases, \href{https://doi.org/10.1103/PhysRevA.87.031602}{Phys. Rev. A {\bf 87}, 031602 (2013)}.

\bibitem{zheng2014fflo} Z. Zheng, M. Gong, Y. Zhang, X. Zou, C. Zhang, and G. Guo, FFLO superfluids in 2D spin-orbit coupled Fermi gases, \href{https://www.nature.com/articles/srep06535}{Sci. Rep. {\bf 4}, 6535 (2014)}.

\bibitem{zhou2011topological} J. Zhou, W. Zhang, and W. Yi, Topological superfluid in a trapped two-dimensional polarized Fermi gas with spin-orbit coupling, \href{https://doi.org/10.1103/PhysRevA.84.063603}{Phys. Rev. A {\bf 84}, 063603 (2011)}.

\bibitem{wu2013unconventional} F. Wu, G.-C. Guo, W. Zhang, and W. Yi, Unconventional Fulde-Ferrell-Larkin-Ovchinnikov pairing states in a Fermi gas with spin-orbit coupling, \href{https://doi.org/10.1103/PhysRevA.88.043614}{Phys. Rev. A {\bf 88}, 043614 (2013)}.

\bibitem{chen2013inhomogeneous} C. Chen, Inhomogeneous Topological Superfluidity in One-Dimensional Spin-Orbit-Coupled Fermi Gases, \href{https://doi.org/10.1103/PhysRevLett.111.235302}{Phys. Rev. Lett. {\bf 111}, 235302 (2013)}.

\bibitem{liu2013topological} X.-J. Liu and H. Hu, Topological Fulde-Ferrell superfluid in spin-orbit-coupled atomic Fermi gases, \href{https://doi.org/10.1103/PhysRevA.88.023622}{Phys. Rev. A {\bf 88}, 023622 (2013)}.

\bibitem{xu2014anisotropic} Y. Xu, R.-L. Chu, and C. Zhang, Anisotropic Weyl Fermions from the Quasiparticle Excitation Spectrum of a 3D Fulde-Ferrell Superfluid, \href{https://doi.org/10.1103/PhysRevLett.112.136402}{Phys. Rev. Lett. {\bf 112}, 136402 (2014)}.

\bibitem{cao2014gapless} Y. Cao, S.-H. Zou, X.-J. Liu, S. Yi, G.-L. Long, and H. Hu, Gapless Topological Fulde-Ferrell Superfluidity in Spin-Orbit Coupled Fermi Gases, \href{https://doi.org/10.1103/PhysRevLett.113.115302}{Phys. Rev. Lett. {\bf 113}, 115302 (2014)}.

\bibitem{iskin2013topological} M. Iskin and A. L. Subasi, Topological superfluid phases of an atomic Fermi gas with in- and out-of-plane Zeeman fields and equal Rashba-Dresselhaus spin-orbit coupling, \href{https://doi.org/10.1103/PhysRevA.87.063627}{Phys. Rev. A {\bf 87}, 063627 (2013)}.

\bibitem{zhou2013exotic} X.-F. Zhou, G.-C. Guo, W. Zhang, and W. Yi, Exotic pairing states in a Fermi gas with three-dimensional spin-orbit coupling, \href{https://doi.org/10.1103/PhysRevA.87.063606}{Phys. Rev. A {\bf 87}, 063606 (2013)}.

\bibitem{dong2013fulde} L. Dong, L. Jiang, and H. Pu, Fulde-Ferrell pairing instability in spin-orbit coupled Fermi gas, \href{https://iopscience.iop.org/article/10.1088/1367-2630/15/7/075014/meta}{New J. Phys. {\bf 15}, 075014 (2013)}.

\bibitem{yi2012molecule} W. Yi and W. Zhang, Molecule and Polaron in a Highly Polarized Two-Dimensional Fermi Gas with Spin-Orbit Coupling, \href{https://doi.org/10.1103/PhysRevLett.109.140402}{Phys. Rev. Lett. {\bf 109}, 140402 (2012)}.

\bibitem{zheng2020cavity} Z. Zheng and Z. D. Wang, Cavity-induced Fulde-Ferrell-Larkin-Ovchinnikov superfluids of ultracold Fermi gases, \href{https://doi.org/10.1103/PhysRevA.101.023612}{Phys. Rev. A {\bf 101}, 023612 (2020)}.

\bibitem{morales2024resonance} L. Morales-Molina and F. Isaule, Resonance detuning mechanism for directing particle flow in ring lattices, \href{https://doi.org/10.1103/PhysRevA.110.053313}{Phys. Rev. A {\bf 110}, 053313 (2024)}.

\bibitem{chen2014three} J. Chen, H. Hu, and X. Gao, Three-component topological superfluid in one-dimensional Fermi gases with spin-orbit coupling, \href{https://doi.org/10.1103/PhysRevA.90.023619}{Phys. Rev. A {\bf 90}, 023619 (2014)}.

\bibitem{qin2015three} F. Qin, F. Wu, W. Zhang, W. Yi, and G.-C. Guo, Three-component Fulde-Ferrell superfluids in a two-dimensional Fermi gas with spin-orbit coupling, \href{https://doi.org/10.1103/PhysRevA.92.023604}{Phys. Rev. A {\bf 92}, 023604 (2015)}.

\bibitem{qiu2016universal} X. Qiu, X. Cui, and W. Yi, Universal trimers emerging from a spin-orbit-coupled Fermi sea, \href{https://doi.org/10.1103/PhysRevA.94.051604}{Phys. Rev. A {\bf 94}, 051604 (2016)}.

\bibitem{zheng2024synthetic} Z. Zheng, Y.-Q. Zhu, S. Zhang, S.-L. Zhu, and Z. D. Wang, Synthetic spin-orbit coupling for the multispin models in optical lattices, \href{https://doi.org/10.1103/PhysRevA.110.033327}{Phys. Rev. A {\bf 110}, 033327 (2024)}.

\bibitem{hu2011probing} H. Hu, L. Jiang, X.-J. Liu, and H. Pu, Probing Anisotropic Superfluidity in Atomic Fermi Gases with Rashba Spin-Orbit Coupling, \href{https://doi.org/10.1103/PhysRevLett.107.195304}{Phys. Rev. Lett. {\bf 107}, 195304 (2011)}.

\bibitem{he2012bcs} L. He and X.-G. Huang, BCS-BEC Crossover in 2D Fermi Gases with Rashba Spin-Orbit Coupling, \href{https://doi.org/10.1103/PhysRevLett.108.145302}{Phys. Rev. Lett. {\bf 108}, 145302 (2012)}.

\bibitem{yu2011spin} Z.-Q. Yu and H. Zhai, Spin-Orbit Coupled Fermi Gases across a Feshbach Resonance, \href{https://doi.org/10.1103/PhysRevLett.107.195305}{Phys. Rev. Lett. {\bf 107}, 195305 (2011)}.

\bibitem{gong2011bcs} M. Gong, S. Tewari, and C. Zhang, BCS-BEC Crossover and Topological Phase Transition in 3D Spin-Orbit Coupled Degenerate Fermi Gases, \href{https://doi.org/10.1103/PhysRevLett.107.195303}{Phys. Rev. Lett. {\bf 107}, 195303 (2011)}.

\bibitem{chen2012bcs} G. Chen, M. Gong, and C. Zhang, BCS-BEC crossover in spin-orbit-coupled two-dimensional Fermi gases, \href{https://doi.org/10.1103/PhysRevA.85.013601}{Phys. Rev. A {\bf 85}, 013601 (2012)}.

\bibitem{zhou2014three} L. Zhou, X. Cui, and W. Yi, Three-Component Ultracold Fermi Gases with Spin-Orbit Coupling, \href{https://doi.org/10.1103/PhysRevLett.112.195301}{Phys. Rev. Lett. {\bf 112}, 195301 (2014)}.

\bibitem{sheehy2015fulde} L. Sheehy, Daniel E, Fulde-Ferrell-Larkin-Ovchinnikov state of two-dimensional imbalanced Fermi gases, \href{https://journals.aps.org/pra/abstract/10.1103/PhysRevA.92.053631}{Phys. Rev. A {\bf 92}, 053631 (2015)}.

\bibitem{kinnunen2018fulde} Kinnunen, Jami J and Baarsma, Jildou E and Martikainen, Jani-Petri and T{\"o}rm{\"a}, P{\"a}ivi, The Fulde--Ferrell--Larkin--Ovchinnikov state for ultracold fermions in lattice and harmonic potentials: a review, \href{https://journals.aps.org/pra/abstract/10.1103/PhysRevA.92.053631}{Rep. Prog. Phys. {\bf 81}, 046401 (2018)}.

\bibitem{he2006loff} He, Lianyi and Jin, Meng and Zhuang, Pengfei, LOFF pairing vs breached pairing in asymmetric fermion superfluids, \href{https://journals.aps.org/pra/abstract/10.1103/PhysRevA.92.053631}{Phys. Rev. B {\bf 73}, 214527(2006)}.
\end{thebibliography}

\end{document}